\documentclass[conference]{IEEEtran}
\usepackage{amsmath, amssymb, bm, cite, epsfig, psfrag}
\usepackage{graphicx}
\usepackage{soul} 
\usepackage[margin=0.60in]{geometry}
\usepackage{dblfloatfix}
\usepackage{array}
\usepackage{lipsum}

\newcolumntype{?}{!{\vrule width 2pt}}

\newcolumntype{P}[1]{>{\centering\hspace{0pt}}p{#1}}
\newcolumntype{M}[1]{>{\centering\hspace{0pt}}m{#1}}
\newcolumntype{L}{>{\centering\arraybackslash}m{3cm}}
\usepackage[font=small]{caption}
\usepackage{epstopdf}	
\usepackage{longtable}
\usepackage{supertabular,booktabs}
\usepackage{bbm}
\usepackage{multirow}
\usepackage[usenames,dvipsnames]{xcolor}

\usepackage{subfigure}
\usepackage{etoolbox}
\usepackage{pbox}
\usepackage{fixltx2e}
\usepackage{tabu}	
\usepackage{filecontents}
\usepackage{enumerate}
\usepackage{textcomp}

\usepackage{colortbl}
\usepackage{fancyhdr}

\usepackage{bm}

\newtoggle{conference}

\togglefalse{conference} 
\graphicspath{{figures/}}

\newcolumntype{?}{!{\vrule width 2pt}}

\fancypagestyle{firststyle}{
	\fancyhf{}
	\fancyhead[L]{Y. Xing and T. S. Rappaport, ``Propagation Measurements and Path Loss Models for sub-THz in Urban Microcells,''  2021 \textit{IEEE International Conference on Communications}, June 2021, pp. 1-6.}     

}
\IEEEoverridecommandlockouts

\begin{document}
\title{Propagation Measurements and Path Loss Models for sub-THz in Urban Microcells}
\author{\IEEEauthorblockN{Yunchou Xing and Theodore S. Rappaport}
	
	\IEEEauthorblockA{	\small NYU WIRELESS, NYU Tandon School of Engineering, Brooklyn, NY, 11201, \{ychou, tsr\}@nyu.edu
	}
\vspace{-0.7cm}
	
	\thanks{This research is supported by the NYU WIRELESS Industrial Affiliates Program and National Science Foundation (NSF) Research Grants: 1909206 and 2037845.}
}

\maketitle
\thispagestyle{firststyle}

\begin{abstract} 
Terahertz frequency bands will likely be used for the next-generation wireless communication systems to provide data rates of hundreds of Gbps or even Tbps because of the wide swaths of unused and unexplored spectrum. This paper presents two outdoor wideband measurement campaigns in downtown Brooklyn (urban microcell environment) in the sub-THz band of 140 GHz with TX-RX separation distance up to 117.4 m: i)  terrestrial urban microcell measurement campaign, and ii) rooftop surrogate satellite and backhaul measurement campaign. Outdoor omnidirectional and directional path loss models for both line-of-sight and non-line-of-sight scenarios, as well as foliage loss (signal attenuation through foliage), are provided at 140 GHz for urban microcell environments. These measurements and models provide an understanding of both the outdoor terrestrial (e.g., 6G cellular and backhaul) and non-terrestrial (e.g., satellite and unmanned aerial vehicle communications) wireless channels, and prove the feasibility of using THz frequency bands for outdoor fixed and mobile cellular communications. This paper can be used for future outdoor wireless system design at frequencies above 100 GHz.
\end{abstract}


\section{Introduction}~\label{sec:intro}
One of the most prominent advancements in the fifth generation (5G) of mobile communications, over prior generations (e.g., 4G LTE), is the use of much wider bandwidth at millimeter wave (mmWave, 30-300 GHz) in comparison to the limited spectrum available at sub-6 GHz frequencies \cite{rappaport19access, viswanathan20A, ghosh195g}. The vast bandwidth enables multi-Gbps data rate operations at mobile devices and various new applications like wireless cognition and centimeter-level positioning \cite{rappaport19access,Kanhere20a,ojas2021ICC}. Both handset terminals and base stations will use highly directional antenna arrays, resulting in huge differences in antenna beamforming and adaptation to wireless channels at mmWave compared to sub-6 GHz frequencies (e.g., narrower antenna beamwidth, higher penetration loss, stronger reflections but much lossier diffractions) \cite{rappaport19access,Sun14b,ghosh195g}. Extensive research has been conducted at frequencies below 100 GHz and several channel models have been developed by standards and different research organizations such as 3GPP, 5GCM, NYUWIRELESS, METIS, and mmMAGIC, which has helped to facilitate the deployment of 5G networks  \cite{3GPP.38.901,A5GCM15,METIS15a,mmMAGIC17a,Sun18a,Rap17a}. 

Terahertz (THz) frequency bands (e.g., frequencies from 100 GHz to 3 THz) are promising bands for the next generation of wireless communications (6G). However, there are notable challenges seen for frequencies above 100 GHz (e.g., high phase noise and Doppler shift, limited output power, and more directional beams), which makes communications in THz bands more challenging \cite{rappaport19access,xing19GC}. Propagation measurements at THz frequency bands are imperative to provide knowledge and understanding of the wireless channels above 100 GHz. 

Most of the measurements at THz bands are focused on short-range indoor scenarios \cite{pometcu18EuCAP,jacob09EUCAP,guan19TVT,nguyen2018comparing}, which are limited by the dynamic range and the cable connected between the transmitter (TX) and receiver (RX) of the vector network analyzer (VNA) based system. Work in \cite{Ju20a,Xing21b} presented an indoor wideband measurement campaign at 142 GHz looking at large indoor radio propagation distances up to 40 m. There are very few outdoor measurement campaigns at frequencies above 100 GHz \cite{ma18channel,abbasi20ICC} and they primarily focus on line-of-sight (LOS) propagation using either reflected materials \cite{ma18channel} or a RF-over-fiber extension \cite{abbasi20ICC} of a VNA based system. 

This paper presents two outdoor measurement campaigns at 142 GHz in an urban microcell (UMi) environment including both LOS and non-LOS (NLOS) scenarios in downtown Brooklyn, NY, using a wideband sliding correlation-based channel sounder system \cite{xing18GC,Mac17JSACb}. The measurement system, procedures, and directional and omnidirectional path losses and models of the terrestrial UMi measurements are described in Section \ref{sec:UMi}. Section \ref{sec:Rooftop} presents the rooftop surrogate satellite and backhaul measurements, as well as the foliage loss at 142 GHz, and shows the isolation between surrogate passive satellite sensors and terrestrial terminals, or between mobile links and terrestrial backhaul links at frequencies above 100 GHz (little interference in the same or adjacent bands). Section \ref{sec:conclusion} provides concluding remarks, showing that THz frequencies can provide outdoor coverage up to 117.4 m in an UMi environment for both LOS and NLOS scenarios with handset-type transmit power (1 mW and 39 dB processing gain \cite{xing18GC,Mac17JSACb}) for future mobile communications.

\section{Outdoor 142 GHz Terrestrial Urban Microcell Measurement Campaign}~\label{sec:UMi}
Propagation in the THz bands is difficult due to the severe path loss in the first meter of propagation from the transmitting antenna and large penetration losses caused by obstructions, which lead to fewer multipath components and clusters in an indoor office compared to lower frequencies (e.g., 28 GHz) \cite{xing19GC,Ju20a,Xing21b}. In Fall 2020, outdoor wideband measurements at 142 GHz were conducted in New York University's (NYU) downtown Brooklyn campus, which is a multipath-rich urban environment \cite{xing18GC}. Two measurement campaigns: i) terrestrial UMi measurement campaign, and ii) rooftop surrogate satellite and backhaul measurement campaign were conducted. These measurements can provide valuable knowledge about the outdoor wireless channels in UMi scenarios and non-terrestrial networks at frequencies above 100 GHz, which may be used for future 6G and beyond communications \cite{rappaport19access,Xing21a}.

\subsection{Wideband 142 GHz Measurement system}~\label{sec:system}
Outdoor wireless channels were measured by transmitting and receiving an 11th order pseudorandom m-sequence centered at 142 GHz with a broadband 1 GHz null-to-null RF bandwidth. The baseband 500 MHz spread spectrum signal was first mixed with an intermediate frequency (IF) of 7 GHz, and was then upconverted with a driving local oscillator (LO) frequency of 135 GHz, resulting in the RF center frequency of 142 GHz. Both the TX and RX used identical mechanically steerable 27 dBi gain horn antennas with 8\textdegree~half power beamwidths (HPBW) in both the azimuth and elevation planes. The TX transmit power was 0 dBm applied by the Virginia Diodes, Inc (VDI) upconverter (27 dBm effective isotropic radiated power, EIRP), and the maximum measurable path loss range of the channel sounder system was 152 dB, operating at a 5 dB SNR threshold \cite{xing18GC,rappaport2015wideband}. A detailed description of the 4th generation channel sounder system at 142 GHz can be found in \cite{xing18GC,Mac17JSACb}. 

\subsection{Terrestrial UMi Measurement Locations and Procedures}
\begin{figure}    
	\centering
	\includegraphics[width=0.450\textwidth]{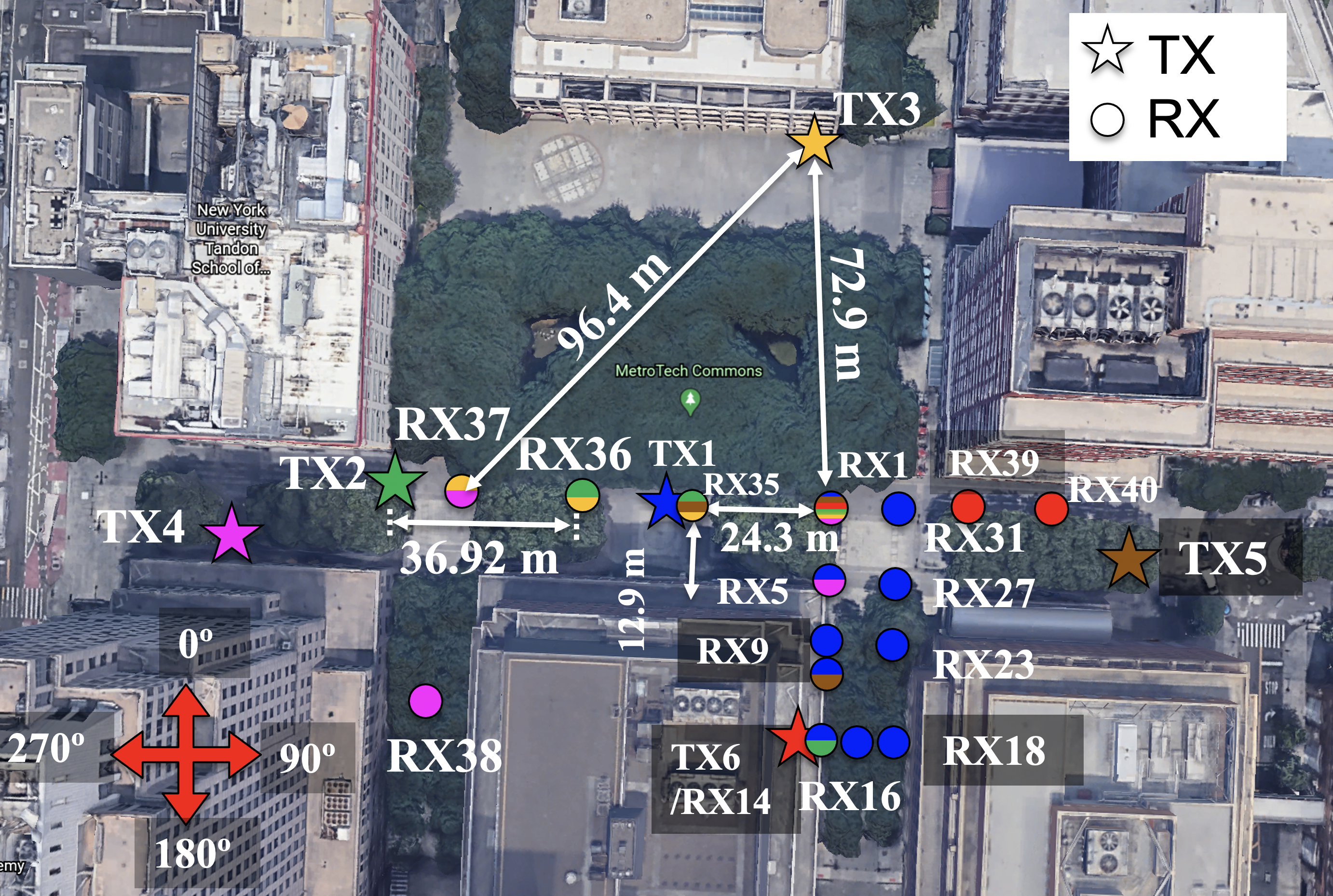}
	\caption{Terrestrial urban microcell measurement campaign in NYU's downtown Brooklyn campus. Six TX locations are identified as stars with different colors and the corresponding RX locations are identified as the same color circles. }
	\label{fig:Mea1Loc}
	\vspace{-0.5cm}
\end{figure}

The terrestrial UMi measurement campaign was designed to study the wireless channels and propagation characteristics in the sub-THz bands for 6G cellular communications in urban microcell and small-cell scenarios. Fig. \ref{fig:Mea1Loc} shows a map of six TX locations and  17 RX locations (with some RX locations reused for more than one TX locations, such as RX1) around NYU's downtown Brooklyn campus. TX1 serves nine RX locations, where five of them are LOS (RX1, RX5, RX 23, RX 27, and RX31) and four of them are NLOS (RX9, RX14, RX16, and RX18). TX2 serves four RX locations, where three of them are LOS (RX1, RX35, and RX36) and one of them is NLOS (RX14). TX3 serves four RX locations, where three of them are LOS (RX35, RX36, and RX37) and the other one is NLOS (RX1). TX4 serves four RX locations, where two of them are LOS (RX3 and RX37) and the other two are NLOS (RX1 and RX38). TX5 serves four RX locations, where two of them are LOS (RX3 and RX35) and the other two are NLOS (RX1 and RX10). TX6 serves three RX locations, where one of them is LOS (RX1) and two of them are NLOS (RX39 and RX40). In total, there are 16 LOS TX-RX location combinations and 12 NLOS TX-RX location combinations with TX-RX separation distances up to 117.4 m. All the 16 LOS locations and 11 out of 12 NLOS locations could successfully receive a signal and measure a power delay profile (PDP) through the channel. 

During the measurements, TXs were set at heights of 4 m above the ground (similar height as lampposts) to emulate small-cell base stations (BS), and the RXs were set at heights of 1.5 m above the ground to emulate mobile user receivers. In the surrounding environment, there are metal lampposts, concrete building walls, paved roads, trees, pedestrians, bare soil ground, concrete pillars, glass windows, vehicles, and glass doors. 

At the beginning and the end of each measurement day, routine calibrations \cite{xing18VTC} were conducted to make sure all the measurements were valid and accurate \cite{rappaport2015wideband,Mac17JSACb,rappaport2013millimeter}. For each TX-RX combination, two elevation angles were used at the BS TX (the TX best pointing elevation angle that the maximum power is received at the RX, and TX antenna downtilt 8\textdegree~from the best TX pointing elevation angle) and three elevation angles were used at each RX (the best RX pointing elevation angle, and RX antenna uptilt and downtilt 8\textdegree~from the best RX pointing elevation angle). For each TX and RX elevation angle combination, the antennas at both the TX and RX were exhaustively rotated by 8\textdegree~HPBW in the azimuth plane (e.g., 45 rotations to cover the 360\textdegree~plane in azimuth) to capture all the possible multipaths in any azimuth directions. A PDP was recorded for each and every unique TX and RX pointing angle, and omnidirectional PDPs were synthesized as introduced in \cite{Sun15a}. Each measured PDP consisted of an average of 20 consecutive instantaneous PDPs to reduce the noise floor, and the averaging factor can be increased at the expense of longer recording time \cite{Rap02a}. 

\subsection{142 GHz Path Loss Models for Terrestrial UMi Measurements}

The 1 m close-in (CI) free space reference distance path loss model \eqref{equ:CI} \cite {Mac15b, Sun16b,rappaport2015wideband} is one of the most commonly used large-scale path loss models to predict the signal strength over distance for various frequencies \cite {Mac15b, Sun16b,3GPP.38.901}:
\begin{equation}
\label{equ:CI}
\small
\begin{split}
PL^{CI}(f_c,d_{\text{3D}})\;\text{[dB]} &= \text{FSPL}(f_c, 1 m) +10n\log_{10}\left( \dfrac{d_{3D}}{d_{0}} \right)+ \chi_{\sigma},\\
\text{FSPL}(f_c,1 m) &= 32.4 + 20\log_{10}(\dfrac{f_c}{1\;\text{GHz}}),
\end{split}
\end{equation}
where FSPL$(f_c, 1 \;\text{m})$ is the large-scale free space path loss at carrier frequency $f_c$ in GHz at 1 m, $n$ is the path loss exponent (PLE), and $\chi_{\sigma}$ is the large-scale fading in dB (a zero mean Gaussian random variable with a standard deviation $\sigma$ in dB) \cite{rappaport2015wideband,Mac15b, Sun16b}.

\begin{figure}    
	\centering
	\includegraphics[width=0.50\textwidth]{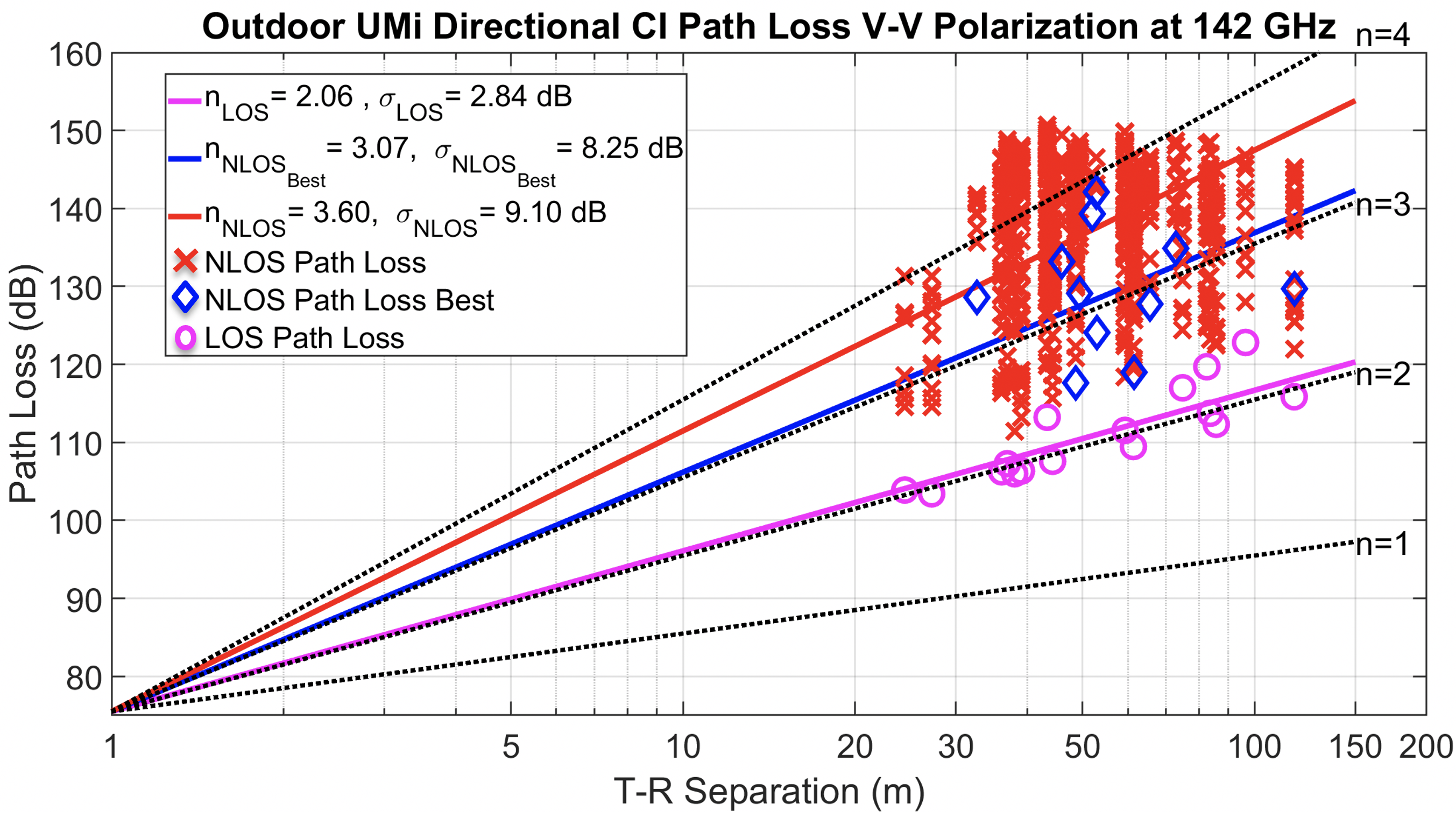}
	\caption{Urban UMi 142 GHz directional path loss scatter plot and outdoor directional CI ($d_0$= 1 m) path loss model for both LOS and NLOS scenarios using 27 dBi gain and 8\textdegree~HPBW horn antennas at both the TX and RX (without antenna gains included for path loss calculations). Each green circle represents the LOS path loss at a LOS location, red crosses represent NLOS path losses measured at arbitrary antenna pointing angles between the TX and RX for NLOS scenarios, and each blue diamond represents the best antenna pointing angles of both the TX and RX to receive the maximum power at the RX for each NLOS TX-RX location combination.}
	\label{fig:directPL}
	\vspace{-0.5cm}
\end{figure}

Fig. \ref{fig:directPL} shows the directional path loss scatter plot and the best-fit directional CI path loss model \eqref{equ:CI} at 142 GHz for both LOS and NLOS scenarios, with antenna gains removed. Each data point in the plot represents one of the directional PDPs for each TX-RX pointing combination. The best-fit minimum mean square error (MMSE) LOS PLE is $n=$ 2.1 with a shadow fading standard deviation of $\sigma =$ 2.8 dB at 142 GHz, showing that the LOS directional channel is only very slightly lossier than the free space propagation ($n=$ 2.0), this likely due to foliage attenuation \cite{Deng15b} or antenna misalignment between the TX and RX. The LOS directional PLE  $n=$ 2.1  at 142 GHz is very close to the LOS directional PLEs of $n=$ 1.9 and 2.0 with shadow fading standard deviations of $\sigma =$  1.1 dB and 1.9 dB in the identical urban microcell environment at 28 and 73 GHz \cite{maccartney19a,rappaport2015wideband,Sun14a}, respectively. The slightly larger LOS PLEs at higher frequencies are likely due to the narrower beam antennas (HPBW= 29\textdegree, 15\textdegree, and 8\textdegree~antennas at 28, 73, and 142 GHz, respectively), since the wider beam antenna will capture more multipath components (e.g., reflections) within the LOS boresight beam and the narrower beam antennas are harder to perfectly align.

The best-fit MMSE $NLOS_{\text{Best}}$ directional PLEs (when the TX and RX antennas are pointing in the direction where the maximum power is received at the RX) and $NLOS_{\text{Arbitrary}}$ directional PLEs are $n=$ 3.1 and 3.6, with shadow fading standard deviations of $\sigma =$ 8.3 dB and 9.1 dB, respectively. The smaller PLE of $NLOS_{\text{Best}}$ directional case compared to the PLE of $NLOS_{\text{Arbitrary}}$ directional situations indicates that there is usually a dominant path (usually a strong reflection) that contains more power than the other multipaths in NLOS directional scenarios. The $NLOS_{\text{Best}}$ and $NLOS_{\text{Arbitrary}}$ directional PLEs $n$ are (3.5 and 4.1) at 28 GHz and (3.1 and 4.6) at 73 GHz \cite{maccartney19a,rappaport2015wideband,Sun14a}. The smaller $NLOS_{\text{Best}}$ PLE $n =$ 3.1 at both 73 and 142 GHz compared to the $NLOS_{\text{Best}}$ PLE $n =$ 3.5 at 28 GHz indicates that reflected paths are stronger at higher frequencies, which gives the surprising, non-intuitive result that there is \textit{less} path loss at higher frequencies when using the best NLOS beam directions. This was also observed in an indoor office \cite{xing19GC,Xing21b}.    

Due to the limitation of the measurable dynamic range of the 142 GHz channel sounder system (152 dB) \cite{xing18GC}, the signals with path loss exceeding 152 dB were not detectable. Therefore, the 142 GHz PLE of NLOS arbitrary ($n=$ 3.60) may not be accurate, as it is conditioned upon reception of a signal (see Fig. \ref{fig:directPL}).

During the measurements, we observed that metal lampposts, concrete building walls, and tinted glass were good reflectors at 142 GHz (providing only 2-8 dB loss to the reflected multipath), which could provide first-order and second-order reflections. However, the partition loss of the outdoor building materials (e.g., penetration loss) at 142 GHz was very large such that concrete walls and tinted windows usually provided 30-50 dB penetration loss.

\begin{figure}
	\centering
	\includegraphics[width=0.5\textwidth]{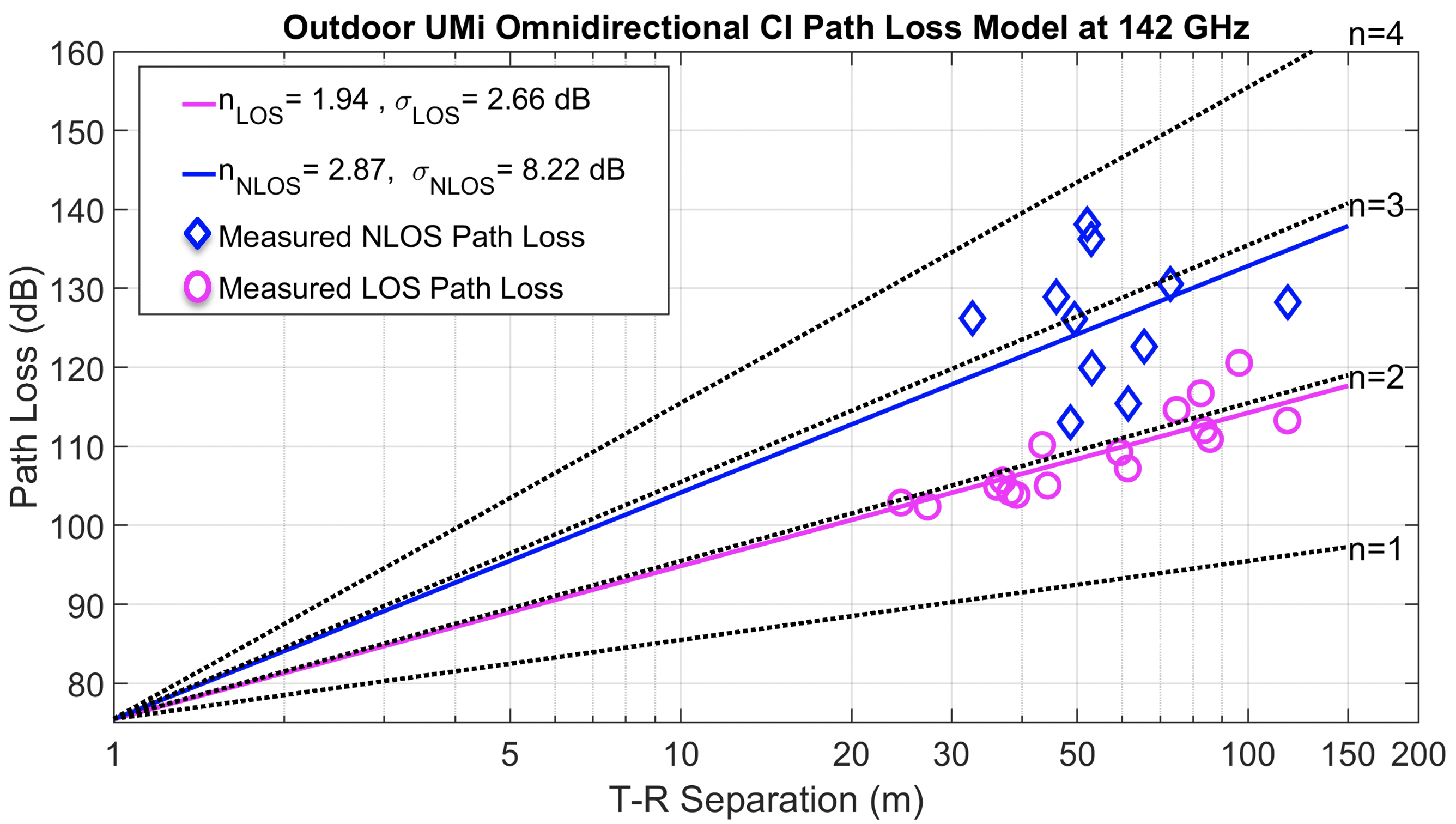}
	\caption{Urban UMi best-fit omnidirectional CI path loss model (without antenna gains) at 142 GHz for both LOS and NLOS situations. The blue diamonds represent the measured omnidirectional path loss at 142 GHz in the NLOS environment and the green circles, conversely, represent the LOS situation.}
	\label{fig:omniPL}
	\vspace{-0.5cm}
\end{figure}

Omnidirectional antenna pattern and omnidirectional received power were synthesized by summing the received powers from every measured non-overlapping directional HPBW antenna pointing angle combination, as described in \cite{Sun15a}. Fig. \ref{fig:omniPL} shows the best-fit omnidirectional CI path loss model and the scatter plot of synthesized omnidirectional measured path loss (without antenna gains) at 142 GHz at NYU courtyard in downtown Brooklyn, NY. 

The LOS omnidirectional PLE at 142 GHz is $n=$ 1.9 with a shadow fading standard deviation $\sigma =$ 2.7 dB, which is slightly lower than the LOS directional PLE of $n=$ 2.1, showing that the omnidirectional antennas would capture slightly more power from all the directions than using boresight-aligned directional antennas, and the boresight LOS multipath component is clearly dominant compared to all other multipaths (e.g., reflected or scattered) in the LOS scenarios. At lower frequencies, the LOS omnidirectional PLEs are nearly identical with $n=$ 2.1 and 1.9 with shadow fading standard deviations of $\sigma =$  3.6 dB and 1.7 dB at 28 and 73 GHz \cite{maccartney19a,rappaport2015wideband,Sun14a}, respectively. 

The 142 GHz NLOS omnidirectional PLE is $n=$ 2.9 with a shadow fading standard deviation $\sigma =$ 8.2 dB which is smaller than but close to the $\text{NLOS}_{\text{Best}}$ directional PLE of $n=$ 3.1 at 142 GHz, indicating that the $NLOS_{Best}$ propagation path is generally the single dominant propagation path among all the multipaths in NLOS scenarios. Thus, accurate beamforming algorithms will be needed to find, capture, and combine the most dominant multipath energy to maintain and extend the outdoor NLOS communication range at frequencies above 100 GHz \cite{Sun14a,Sun14b}. The NLOS omnidirectional PLEs are larger with more variability at lower frequencies, such that $n=$ 3.4 and 2.8 with shadow fading standard deviations of $\sigma =$  9.7 dB and 8.7 dB at 28 and 73 GHz, respectively. Future data processing will determine the relative strength of the strongest few multipath components in THz NLOS channels. Comparing path loss models between different frequencies, it is clear that reflections are stronger at 73 and 142 GHz compared to at 28 GHz, and THz channels of 142 GHz are very similar to 73 GHz except for the path loss in the first meter of propagation when energy spreads into the far field \cite{Xing21b}.

\begin{figure}
	\centering
	\includegraphics[width=0.4\textwidth]{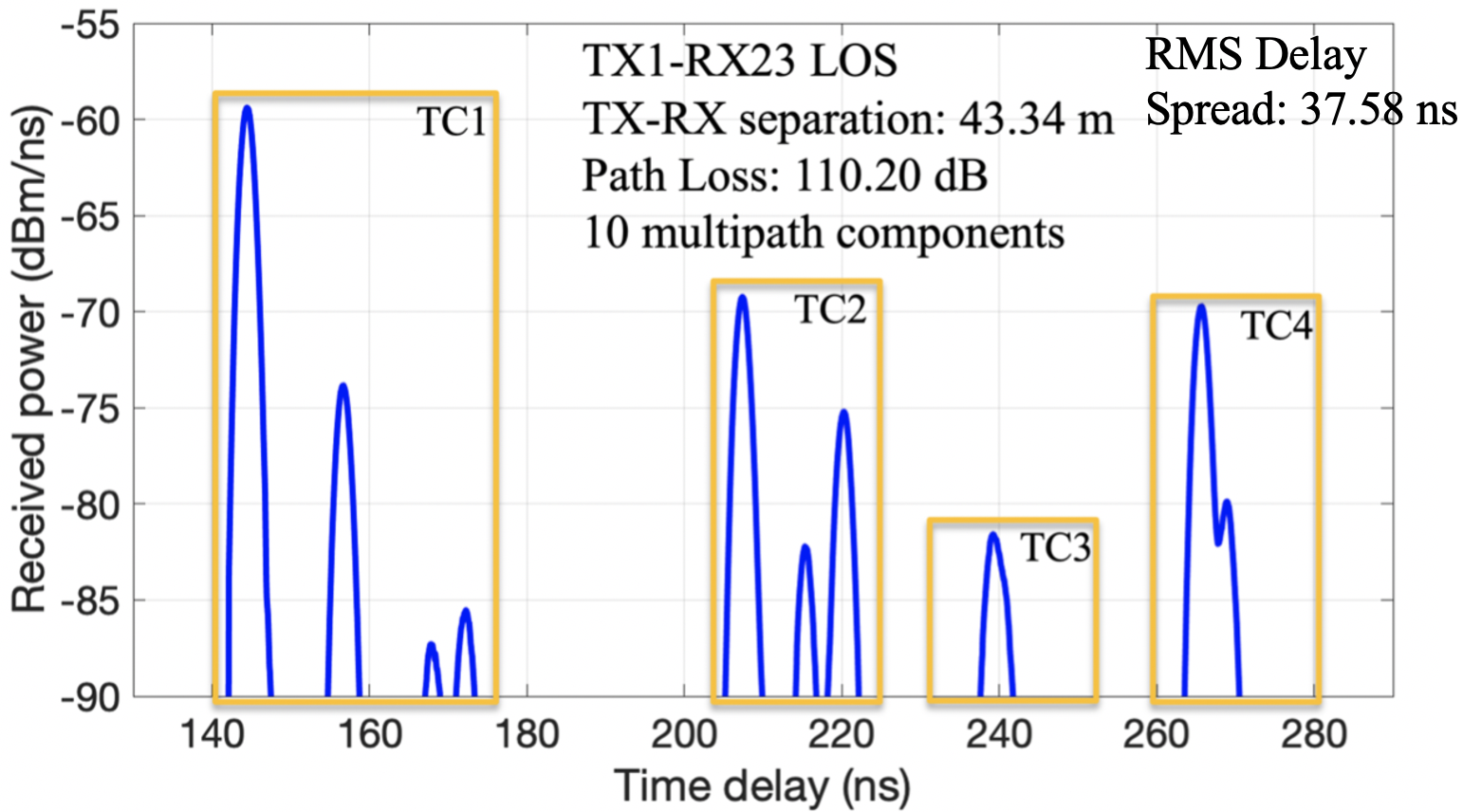}
	\caption{Omnidirectional PDP of TX1-RX23 which is in the LOS scenario with a separation distance of 43.34 m. There are 10 multipath components in four time clusters (TC), where the first multipath component is the LOS path, and the others are reflected paths from the neighbor buildings and surrounding lampposts as shown in Fig. \ref{fig:Mea1Loc}.}
	\label{fig:TX1RX23}
	\vspace{-0.5cm}
\end{figure}

A typical omnidirectional PDP measured between TX1 and RX23 in the LOS scenario with a TX-RX separation distance of 43.34 m is shown in Fig.\ref{fig:TX1RX23} (omnidirectional PDPs in NLOS scenarios are being synthesized with the help of the NYURay ray-tracer \cite{ojas2021ICC}, and will be available in future work). Fig. \ref{fig:TX1RX23} shows four time clusters \cite{Samimi2016MTT}, where there are between 1 to 4 multipath components in each time cluster and 10 multipath components in total. The first multipath component is the boresight LOS path from TX1 to RX23 with a propagation time delay of 144.47 ns. The other multipath components are either reflected from the neighboring buildings, pillars, or surrounding lampposts, resulting in a root mean square delay spread of 37.6 ns.

Conventional wisdom before this work is that LOS communication would be the main use case (e.g., 6G cellular and backhaul) at 142 GHz in outdoor environments. However, we observed that many outdoor construction materials (e.g., metal lamppost, concrete building walls, and tinted glass) served as excellent reflectors at 142 GHz which enabled good NLOS coverage up to 117.4 m, as shown in Figs. \ref{fig:directPL} and \ref{fig:omniPL}. 

\section{Rooftop surrogate satellite and backhaul measurement campaign}~\label{sec:Rooftop}
\begin{figure}    
	\centering
	\includegraphics[width=0.50\textwidth]{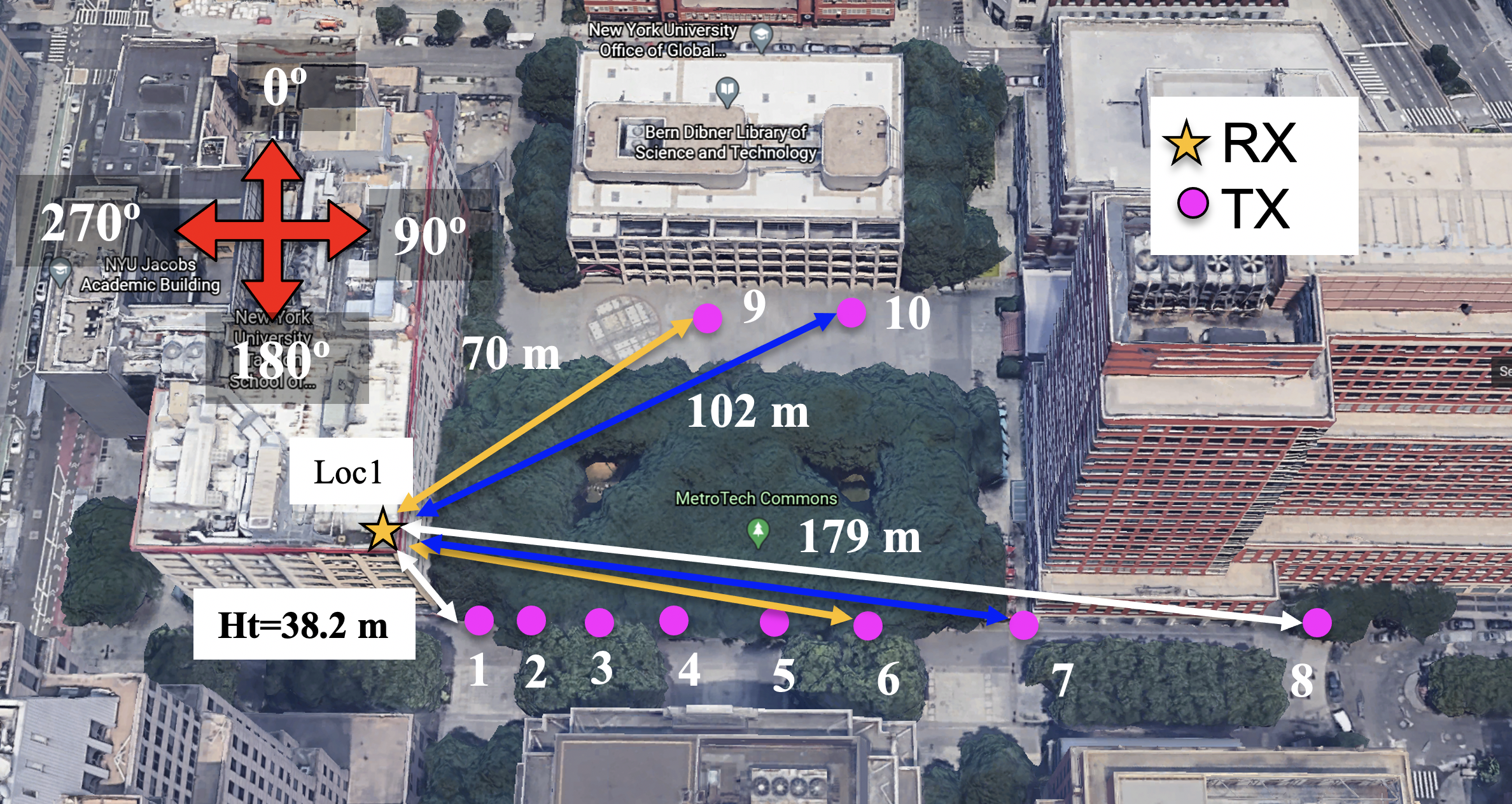}
	\caption{Rooftop surrogate satellite and backhaul measurement campaign at NYU courtyard in downtown Brooklyn, NY. The surrogate satellite (and backhaul) receiver RX location is 38.2 m above the ground on the rooftop identified as a yellow star. Ten mobile TX locations on the ground are identified as purple circles. The LOS elevation pointing angles from TX1-8 to the RX location are 80\textdegree~to 15\textdegree, respectively.}
	\label{fig:Mea2Loc}
	\vspace{-0.5cm}
\end{figure}

The second measurement campaign used a rooftop surrogate satellite and backhaul RX for non-terrestrial network scenarios (e.g., air-to-ground communications and fixed point-to-point backhaul over the roof) \cite{Xing21a}. This measurement campaign attempted to emulate ground-to-satellite and ground-to-unmanned aerial vehicle (UAV) communications, providing insights into potential issues for ground-to-satellite interference and spectrum sharing and coexistence techniques. This measurement campaign uses the same channel sounder system and antennas as described in Section \ref{sec:system} but with ground mounted TX locations using rotating antennas at 1.5 m height. Fig. \ref{fig:Mea2Loc} presents the rooftop RX and ground TX locations in NYU's downtown Brooklyn campus.

\subsection{Rooftop Surrogate Satellite and Backhaul Measurement Locations and Procedures}
For the rooftop surrogate satellite and backhaul measurement campaign, the RX was placed on the rooftop corner of the 6 MetroTech building which is 38.2 m above ground, emulating a surrogate satellite passive receiver. Horn antennas with 8\textdegree~half power beam width (HPBW) were used at both the rooftop RX and ground-based TXs, where the rooftop RX antenna was mechanically steered and extensively rotated to consider all possible pointing combinations in the search of energy from the ground-based TXs that were systematically rotated over all azimuth directions and several elevation directions described subsequently \cite{Xing21a}. In satellite communications, the received power level from the ground to the satellite will be highly dependent upon the sidelobes of the antennas and earth-based TX elevation angles, due to atmospheric absorption and different slant path lengths \cite{Xing21a}. Thus, eight ground-based TX (1.5 m antenna height) locations (TX1-8) were chosen to have the LOS boresight elevation angles to the roof-mounted RX in 10\textdegree~decrements ranging from 80\textdegree~to 10\textdegree~to study the relationship between received power and elevation angles \cite{Xing21a}. Due to the space limitation of the campus measurement area, the farthest TX provided a 15\textdegree~elevation angle boresight to the RX instead of 10\textdegree. The channel sounder requires a clear LOS link for calibration, but TX locations 1 through 8 were somewhat blocked by tree foliage. To overcome this issue, two additional TX locations - TX 9 and 10 were chosen, which had the same link lengths as TX 6 and 7, respectively, for free space calibration without any link obstructions \cite{Xing21a}.

A multipath PDP for the (sometimes foliage-blocked) LOS boresight TX-RX pointing combination was first measured at each TX location, and then the TX antenna was rotated 360\textdegree~in the azimuth plane by steps of 8\textdegree, and at elevation angles of 0\textdegree, 8\textdegree, 16\textdegree, 24\textdegree, and 32\textdegree. For each TX pointing angle, the RX searched every direction to capture any signals (e.g., direct path, reflected, or scattered rays) \cite{Xing21a}.  

\begin{table*}[]\caption{Foliage-blocked LOS links (TX1-7 to RX) and clear LOS links (TX8-10 to RX) from the ground-mounted TXs (1.5 m ht) to the roof-mounted RX (38.2 m above the ground) at 142 GHz, with TX transmit power of -2 dBm and identical 27 dBi gain horn antennas at both of the TX and RX. The predicted received power (assuming free space propagation) $Pr_{FS}$ in dBm, measured received power through foliage-blocked links $Pr$ in dBm, and corresponding foliage loss $(Pr_{FS}-Pr)$ in dBm at different TX-RX separation distances and elevation angles are presented. The negligible difference of the predicted and measured received power of the clear LOS links at TX8, TX9, and TX10 validated the accuracy of the 142 GHz channel sounder system used in this paper.}\label{tab:foilage1}
	\centering
	\begin{tabular}{|l?c|c|c|c|c|c|c?c|c|c|}
		\hline
		& \multicolumn{7}{c?}{\textbf{Foliage-blocked LOS Links}} & \multicolumn{3}{c|}{\textbf{Clear LOS links}} \\ \hline
		\textbf{Ground-mounted TX Location} & TX1   & TX2 & TX3   & TX4   & TX5   & TX6   & TX7   & TX8           & TX9           & TX10          \\ \hline
		\textbf{TX-RX Separation Distance [m]} & 38.5  & 40.4  & 43.9  & 49.3  & 58.6  & 70.6  & 102.7 & 178.9         & 70.6          & 102.7         \\ \hline
		\textbf{LOS Boresight Elevation $\theta$}    & 80\textdegree    & 70\textdegree  & 60\textdegree    & 50\textdegree    & 40\textdegree    & 30\textdegree    & 20\textdegree    & 15\textdegree            & 30\textdegree            & 20\textdegree            \\ \hline
		\textbf{Predictied Free Space Received Power $Pr_{FS}$ [dBm]}         & -55.2 & -55.6 & -56.3 & -57.3 & -58.8 & -60.5 & -63.7 & -68.5         & -60.5        & -63.7         \\ \hline
		\textbf{Measured Received Power $Pr$ [dBm]}           &-60.2  & -62.0 &-59.3 & -65.8 & -63.4 & -70.8 & -74.5 & -68.6        &   -60.9       &  -64.3        \\ \hline
		\textbf{Foliage Loss $(Pr_{FS} - Pr)$ [dB]}          & 5.0     & 6.4 & 3.0     & 8.5   & 4.6   & 10.3  & 10.8  & 0.1           & 0.4           & 0.6           \\ \hline
		\textbf{Average Foliage Loss and Variance}          & \multicolumn{7}{c?}{6.9 dB, 3.0 dB}                    & \multicolumn{3}{c|}{Free Space Calibrations} \\ \hline
	\end{tabular}
	\vspace{-0.5cm}
\end{table*}

\begin{figure}    
	\centering
	\includegraphics[width=0.50\textwidth]{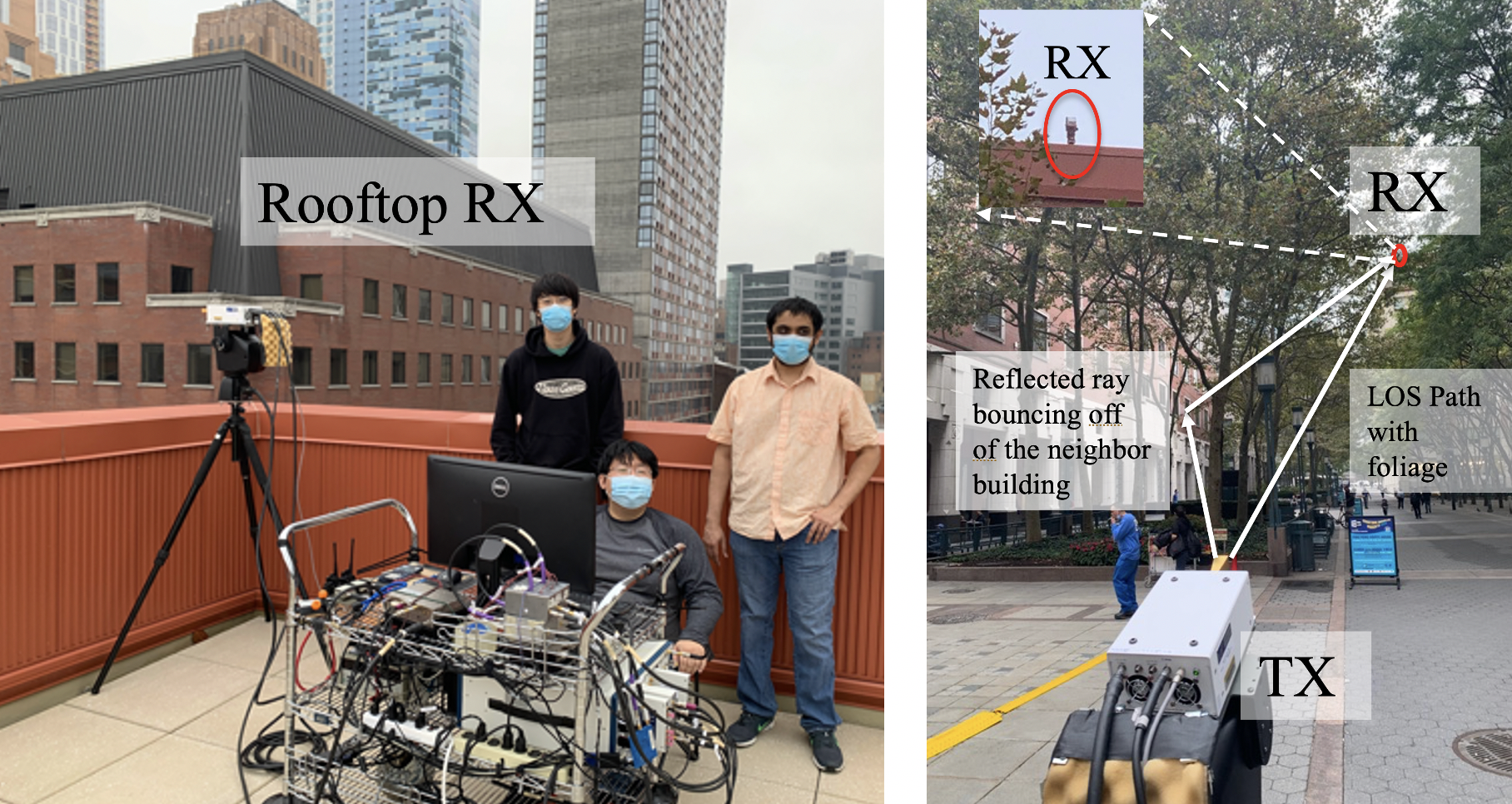}
	\caption{The RX is at heights of 1.5 m above the roof corner, which prevents the RX from being shadowed by the railing boundary, emulating a passive receiver in a satellite. The TXs are at heights of 1.5 m above ground working as mobile terminals.}
	\label{fig:RT1}
	\vspace{-0.5cm}
\end{figure}

\subsection{Rooftop Surrogate Satellite and Backhaul Measurement Results with Foliage Loss Analysis}~\label{sec:results}

\begin{figure}    
	\centering
	\includegraphics[width=0.5\textwidth]{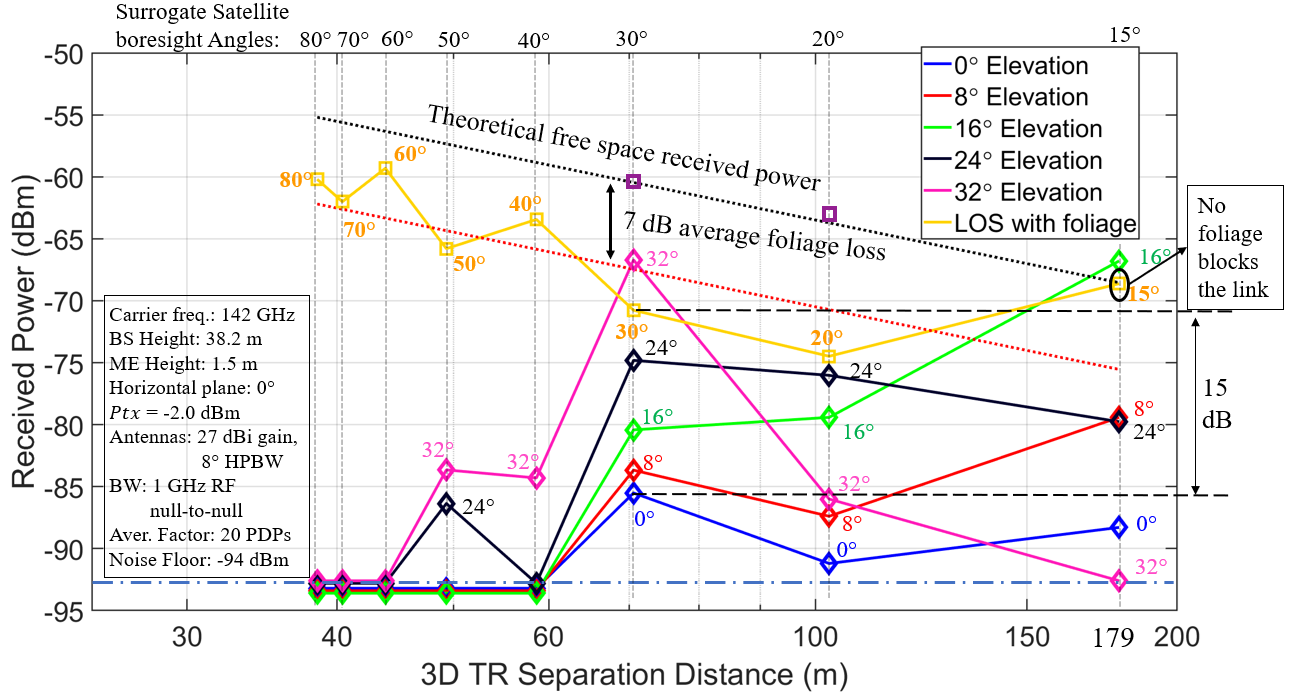}
	\caption{The rooftop base station (38.2 m above the ground) received power vs. different distances and different elevation angles from ground users (1.5 m above the ground) at 142 GHz \cite{Xing21a}. }
	\label{fig:RTPR}
	\vspace{-1.0 em}
\end{figure}

The received power at the roof-mounted RX (38.2 m above the ground) from the ground-mounted TXs (1.5 m above the ground) at different distances and elevation angles are shown in Fig. \ref{fig:RTPR} and Table \ref{tab:foilage1} (0\textdegree~elevation angle signifies the horizontal plane, and the positive values represent the TX elevation boresight angles above the horizon). When the ground-based TX antenna is pointing at a 0\textdegree~elevation angle on the horizon (the blue curve in Fig.\ref{fig:RTPR}), there is virtually no power (other than sidelobe radiation) captured by the rooftop RX even when the ground TX has antenna pattern energy leaking from its main lobe antenna pattern while pointing nearly directly to the roof (e.g., with boresight elevation angle of 15\textdegree). The worst case of interference was found when the TX is at Location 6 (70 m) and Fig.\ref{fig:RTPR} shows how raising the elevation angle of the ground-based transmitter dramatically increases energy detected by the roof-mounted RX, due to antenna pattern leakage and multipath from surrounding buildings.

The heights of birch trees in the NYU courtyard, were 8-10 m as shown in Fig. \ref{fig:RT1}. The foliage of the birch trees was between 5-10 m height above the ground and began falling from the trees during the measurements in November 2020. Slant link length (L) through foliage at different TX-RX separation distances and elevation angles ($\theta$ ranges between 80\textdegree-20\textdegree) is approximately $L= 5~\text{m}/sin(\theta) =$ 5-15 m with an average slant link length of 7.9 m (3.4 m standard deviation) through the foliage. 

The predicted received power (assuming free space propagation) $Pr_{FS}$ in dBm, measured received power through foliage-blocked links $Pr$ in dBm (with the TX transmit power of -2 dBm and identical 27 dBi gain horn antennas at both of the TX and RX), and corresponding foliage loss $(Pr_{FS}-Pr)$ in dBm at different TX-RX separation distances and elevation angles are presented in Table \ref{tab:foilage1}. The negligible difference of the predicted and measured received power of the clear LOS links at TX8, TX9, and TX10 validated the accuracy of the 142 GHz channel sounder system used in this paper. The foliage loss at 142 GHz ranges from 3-11 dB with a 6.9 dB average loss beyond free space and a standard deviation of 3.0 dB, revealing an average foliage attenuation rate of 0.9 dB/m which is higher than the 0.4 dB/m foliage attenuation rate at 73 GHz \cite{Deng15b}, indicating foliage loss (signal attenuation through foliage) increase with carrier frequencies.



\section{Conclusion}\label{sec:conclusion}
Two outdoor radio propagation measurement campaigns in urban microcells at 142 GHz are presented in this paper, using a wideband sliding correlation-based channel sounder with identical narrow-beam 27 dBi gain horn antennas at both the TX and RX. Omnidirectional and directional CI path loss models with a 1 m reference distance are provided for both LOS and NLOS scenarios in the outdoor terrestrial urban microcell environment at 142 GHz. The path loss results are very encouraging and show that in NLOS scenarios there is usually one or a few dominant paths (e.g., the best pointing beam $NLOS_{\text{Best}}$) compared to the other multipath components (e.g., reflected or scattered), and accurate beamforming algorithms will be needed to find, capture, and combine the most dominant multipath energy to maintain and extend the outdoor NLOS communication range at frequencies above 100 GHz. Suprisingly, metal lamppost, concrete walls, and tinted glass perform as good reflectors at 142 GHz (providing only 2-8 dB loss to the reflected multipath) which provide good coverage using only moderate TX power up to 117.4 m in NLOS scenarios. Elevation angles and antenna sidelobes play an important role in the received power in air-to-ground channels. Reflected rays by neighboring buildings from the ground to the air are about 20 dB lower than the direct LOS path, showing that if the propagation of ground terminals is kept on the horizon (e.g., $\leq$ 15\textdegree), with reduced sidelobes there may not be any interference in the same or adjacent bands between surrogate satellites and terrestrial terminals, or between terrestrial backhaul links and mobile links at frequencies above 100 GHz. The average foliage loss in urban microcell environments is 6.9 dB (a 7-8 m slant link length through foliage) with the foliage attenuation rate of 0.9 dB/m at 142 GHz. Measurements and models presented here contribute to the understanding that THz will be useful for urban wireless communication, even in NLOS, as well as non-terrestrial (e.g., satellite and UAV communications) wireless channels, and may help with spectrum sharing techniques between the satellites and the ground terminals.

\bibliographystyle{IEEEtran}
\bibliography{Indoor140GHz}

\end{document}